\documentclass[onecolumn,prd,superscriptaddress,altaffilletter]{revtex4-2}
\usepackage{moreverb}
\usepackage{cprotect}
\usepackage{graphicx}
\usepackage{float}
\usepackage{amsmath}
\usepackage{amssymb}
\usepackage{amsfonts}
\usepackage{comment}
\usepackage{graphicx}
\usepackage{interval}
\usepackage[nolist]{acronym}
\usepackage[usenames,dvipsnames]{xcolor}
\usepackage{xspace}
\usepackage{subfigure}
\usepackage{tikz}
\usepackage[utf8]{inputenc}
\usepackage[T1]{fontenc}
\usepackage{url}
\usepackage[squaren]{SIunits}
\usetikzlibrary{arrows,shapes,trees,decorations.pathreplacing}
\newcommand\BibTeX{{\rmfamily B\kern-.05em \textsc{i\kern-.025em b}\kern-.08em
T\kern-.1667em\lower.7ex\hbox{E}\kern-.125emX}}

\usepackage[colorlinks=true,citecolor=blue,filecolor=blue,linkcolor=blue,urlcolor=blue,pdftex]{hyperref}

\published{27 July, 2022: \href{https://doi.org/10.1002/spe.3120}{Wiley Journal of Software: Practice and Experience}}

\begin{document}

\title{Collaborative Experience between Scientific Software Projects using Agile~Scrum~Development}

\author{A.~L.~Baxter} 
\email[SNEWS corresponding author: ]{adepoian@purdue.edu}
\affiliation{Department of Physics and Astronomy, Purdue University, West Lafayette, IN 47907, USA}

\author{S.~Y.~BenZvi}  
\affiliation{Department of Physics and Astronomy, University of Rochester, Rochester, NY 14627, USA}

\author{W.~Bonivento} 
\affiliation {INFN sezione di Cagliari Istituto Nazionale, Complesso Universitario di Monserrato - S.P. per Sestu Km 0.700, I-09042 Monserrato (Cagliari), Italy}

\author{A.~Brazier} 
\affiliation {Center for Advanced Computing, Cornell University, Ithaca, NY 14853, USA}

\author{M.~Clark} 
\affiliation{Department of Physics and Astronomy, Purdue University, West Lafayette, IN 47907, USA}

\author{A.~Coleiro} 
\affiliation{Université de Paris, CNRS, AstroParticule et Cosmologie, F-75013, Paris, France}

\author{D.~Collom} 
\affiliation{Las Cumbres Observatory, Goleta, CA  93117, USA}

\author{M.~Colomer-Molla} 
\affiliation{Université de Paris, CNRS, AstroParticule et Cosmologie, F-75013, Paris, France}
\affiliation{Instituto de F\'isica Corpuscular (CSIC - Universitat de Val\`encia) c/ Catedr\'atico Jos\'e Beltr\'an, 2 E-46980 Paterna, Valencia, Spain}

\author{B.~Cousins} 
\email[SCiMMA corresponding author: ]{bfc5288@psu.edu}
\affiliation{Institute for Computational and Data Sciences, The Pennsylvania State University, University Park, PA 16802, USA}
\affiliation{Department of Physics, The Pennsylvania State University, University Park, PA 16802, USA}

\author{A.~Delgado~Orellana} 
\affiliation{Pontificia Universidad Catolica de Chile, Santiago, Región Metropolitana, Chile}

\author{D.~Dornic}  
\affiliation{Aix Marseille Univ, CNRS/IN2P3, CPPM, Marseille, France}

\author{V.~Ekimtcov} 
\affiliation{National Center for Supercomputing Applications, University of Illinois at Urbana-Champaign, Urbana, IL 61801, USA}

\author{S.~ElSayed} 
\affiliation{Department of Computer Science, University of California Santa Barbara, CA 93106, USA}

\author{A.~Gallo~Rosso}  
\affiliation{Department of Physics, Laurentian University, Sudbury ON P3E 2C6, Canada}

\author{P.~Godwin} 
\affiliation{Institute for Gravitation and the Cosmos, The Pennsylvania State University, University Park, PA 16802, USA}
\affiliation{Department of Physics, The Pennsylvania State University, University Park, PA 16802, USA}

\author{S.~Griswold}
\affiliation{Department of Physics and Astronomy, University of Rochester, Rochester, NY 14627, USA}

\author{A.~Habig} 
\affiliation{Department of Physics and Astronomy, University of Minnesota Duluth, Duluth, MN, 55812, USA}

\author{R.~Hill} 
\affiliation{Department of Physics, Laurentian University, Sudbury ON P3E 2C6, Canada}

\author{S.~Horiuchi} 
\affiliation{Center for Neutrino Physics, Department of Physics, Virginia Tech, Blacksburg, VA 24061, USA}

\author{D.~A.~Howell} 
\affiliation{Las Cumbres Observatory, Goleta, CA 93117, USA}
\affiliation{University of California Santa Barbara, CA 93106, USA}

\author{M.~W.~G.~Johnson} 
\affiliation{National Center for Supercomputing Applications, University of Illinois at Urbana-Champaign, Urbana, IL 61801, USA}

\author{M.~Jurić} 
\affiliation{DiRAC Institute and the Department of Astronomy, University of Washington, Seattle, WA 98195, USA}

\author{J.~P.~Kneller} 
\affiliation{Department of Physics, NC State University, Raleigh, NC 27695, USA}

\author{A.~Kopec} 
\affiliation{Department of Physics and Astronomy, Purdue University, West Lafayette, IN 47907, USA}

\author{C.~Kopper} 
\affiliation{Department of Physics and Astronomy, Michigan State University, East Lansing, MI 48824, USA}

\author{V.~Kulikovskiy} 
\affiliation{INFN Sezione di Genova, Via Dodecaneso 33, Genova, 16146 Italy}

\author{M.~Lamoureux} 
\affiliation{INFN Sezione di Padova \& Universit\`a di Padova, Dipartimento di Fisica, Padova, Italy}

\author{R.~F.~Lang} 
\affiliation{Department of Physics and Astronomy, Purdue University, West Lafayette, IN 47907, USA}

\author{S.~Li} 
\affiliation{Department of Physics and Astronomy, Purdue University, West Lafayette, IN 47907, USA}

\author{M.~Lincetto}
\affiliation{Aix Marseille Univ, CNRS/IN2P3, CPPM, Marseille, France}

\author{W.~Lindstrom} 
\affiliation{Las Cumbres Observatory, Goleta, CA  93117, USA}

\author{M.~W.~Linvill} 
\affiliation{Department of Physics and Astronomy, Purdue University, West Lafayette, IN 47907, USA}

\author{C.~McCully} 
\affiliation{Las Cumbres Observatory, Goleta, CA  93117, USA}

\author{J.~Migenda} 
\affiliation{Department of Physics, King’s College London, London WC2R 2LS, UK}

\author{D.~Milisavljevic} 
\affiliation{Department of Physics and Astronomy, Purdue University, West Lafayette, IN 47907, USA}

\author{S.~Nelson} 
\affiliation{DiRAC Institute and the Department of Astronomy, University of Washington, Seattle, WA 98195, USA}

\author{R.~Novoseltseva} 
\affiliation{Baksan Neutrino Observatory, Institute for Nuclear Research of Russian Academy of Sciences, Neytrino, Kabardino-Balkarian Republic, Russia, 361609}

\author{E.~O'Sullivan} 
\affiliation{Deptartment of Physics and Astronomy, Uppsala University, Box 516, S-75120 Uppsala, Sweden}

\author{D.~Petravick} 
\affiliation{National Center for Supercomputing Applications, University of Illinois at Urbana-Champaign, Urbana, IL 61801, USA}

\author{B.~W.~Pointon} 
\affiliation{Department of Physics, British Columbia Institute of Technology, Burnaby, BC, V5G 3H2, Canada}
\affiliation{TRIUMF, Vancouver, BC V6T 2A3, Canada}

\author{N.~Raj} 
\affiliation{TRIUMF, Vancouver, BC V6T 2A3, Canada}

\author{A.~Renshaw}  
\affiliation{Department of Physics, University of Houston, Houston, TX 77204, USA}

\author{J.~Rumleskie} 
\affiliation{Department of Physics, Laurentian University, Sudbury ON P3E 2C6, Canada}

\author{T.~Sonley} 
\affiliation{SNOLAB, Sudbury, ON P3Y 1N2, Canada}

\author{R.~Tapia} 
\affiliation{Institute for Computational and Data Sciences, The Pennsylvania State University, University Park, PA 16802, USA}
\affiliation{Department of Physics, The Pennsylvania State University, University Park, PA 16802, USA}

\author{J.~C.~L.~Tseng} 
\affiliation{Department of Physics, University of Oxford, Keble Road, Oxford OX1 3RH, UK}

\author{C.~D.~Tunnell} 
\affiliation{Department of Computer Science, Rice University, 6100 Main St, Houston, TX, 77005, USA}
\affiliation{Department of Physics, Astronomy, Rice University, 6100 Main St, Houston, TX, 77005, USA}

\author{G.~Vannoye} 
\affiliation{Aix Marseille Univ, CNRS/IN2P3, CPPM, Marseille, France}

\author{C.~F.~Vigorito}  
\affiliation{Department of Physics, University of Torino \& INFN, via Pietro Giuria 1, 10125 Torino, Italy}

\author{C.~J.~Virtue} 
\affiliation{Department of Physics, Laurentian University, Sudbury ON P3E 2C6, Canada}
\affiliation{SNOLAB, Sudbury, ON P3Y 1N2, Canada}

\author{C.~Weaver} 
\affiliation{Department of Physics and Astronomy, Michigan State University, East Lansing, MI 48824, USA}

\author{K.~E.~Weil} 
\affiliation{Department of Physics and Astronomy, Purdue University, West Lafayette, IN 47907, USA}

\author{L.~Winslow}
\affiliation{Massachusetts Institute of Technology, Cambridge, MA 02139, USA}

\author{R.~Wolski} 
\affiliation{Department of Computer Science, University of California Santa Barbara, CA 93106, USA}

\author{X.~J.~Xu} 
\affiliation{Université Libre de Bruxelles, 1050 Bruxelles, Belgium}

\author{Y.~Xu} 
\affiliation{Department of Computer Science, Rice University, 6100 Main St, Houston, TX, 77005, USA}

\collaboration{The SCiMMA and SNEWS Collaborations}



\begin{abstract}

Developing sustainable software for the scientific community requires expertise in software engineering and domain science. This can be challenging due to the unique needs of scientific software, the insufficient resources for software engineering practices in the scientific community, and the complexity of developing for evolving scientific contexts. While open-source software can partially address these concerns, it can introduce complicating dependencies and delay development. These issues can be reduced if scientists and software developers collaborate. We present a case study wherein scientists from the SuperNova Early Warning System collaborated with software developers from the Scalable Cyberinfrastructure for Multi-Messenger Astrophysics project. The collaboration addressed the difficulties of open-source software development, but presented additional risks to each team. For the scientists, there was a concern of relying on external systems and lacking control in the development process. For the developers, there was a risk in supporting a user-group while maintaining core development. These issues were mitigated by creating a second Agile Scrum framework in parallel with the developers' ongoing Agile Scrum process. This Agile collaboration promoted communication, ensured that the scientists had an active role in development, and allowed the developers to evaluate and implement the scientists' software requirements. The collaboration provided benefits for each group: the scientists actuated their development by using an existing platform, and the developers utilized the scientists' use-case to improve their systems. This case study suggests that scientists and software developers can avoid scientific computing issues by collaborating and that Agile Scrum methods can address emergent concerns.

\end{abstract}



\maketitle

\section{Introduction}\label{sec:intro}
The creation of scientific computing applications must address several software development challenges unique to the scientific software context. Unlike many commercial applications, a scientific application must often provide a precise solution to a problem before the software is modified or discarded~\cite{Brooks}. This need for precise scientific solutions requires domain knowledge, leading to the inclusion of scientists on development teams. Indeed, scientists spend about 30\% of their time developing scientific software, though they may not have prior training in software engineering and often must learn these methods during the development process itself~\cite{Hannay2009,Merali_nature}. Moreover, budgetary constraints or time limitations often make it impractical to employ professional developers to share responsibility for software integrity with the scientists. The timescales of scientific projects can be decades long with high personnel turnover, producing challenges for software sustainability. In particular, scientists who fulfill a development role can create legacy technical debt when they change roles or projects as they exploit new career opportunities. Efforts to introduce a new career trajectory of ``research software engineer'' attempt to address this problem of software sustainability~\cite{rse}. Another approach to this issue involves the use of community-supported software, namely open-source codebases, that can ``outlive'' the tenure of any one developer~\cite{Wilson_2014,Goble_2014}. Open-source is freely available and community maintained, but the release and maintenance life-cycles are not coordinated between technologies and services. Thus, the maintenance procedures for an application that contains multiple open-source technologies must be able to account for and manage the different software life-cycles of its dependencies, introducing a risk that a required feature may be unavailable by a software dependency.  Moreover, the community-based maintenance model of open-source codebases can introduce issues with the rapid time frame of scientific software development. While users can submit feature requests, issues, and bug fixes in the form of pull-requests to the external codebase, the integration of these requests can take days or weeks, if they are integrated at all~\cite{gousios2014exploratory}.

These complexities are often unaddressed in modern scientific software development contexts since resolving them introduces additional risks to a positive and timely scientific outcome. Simply adding the tooling necessary to develop and maintain long-lived services can slow the time to scientific solution due to the time required to both learn and apply this tooling. As a result, the use of software auditing, life-cycle, and management methods like version control and issue tracking for scientific applications varies over a large range from absent to proficient, with many implementations either budding or unused~\cite{Wilson_2014}. Moreover, service-based scientific applications often consist of deep, bespoke software stacks, developed by domain experts who become integral to their sustainability. The development teams are small and, once in production, may consist of only a single expert developer who knows the system well-enough to provide support for issues. This lack of scalability causes risks for scientific applications that serve large communities of scientists. These issues have led to a ``chasm'' between scientific computing and formal software development~\cite{kelly2007software}. However, it has been suggested that this shortcoming can be overcome by building collaborative bridges between the scientific computing and software development communities~\cite{kelly2007software,storer2017bridging}.

In this paper, we describe the experience of scientific computing developers~\cite{SNEWS:2020tbu} who collaborate with a professional Agile software development team~\cite{Chang:2019edd} to create a scientific application~\cite{patrick_godwin_2021_4437579} that depends on a cloud-based service created independently by the software developers~\cite{hopskotchSite}. This collaboration allowed the scientific team to mitigate the risk of legacy software~\cite{Bennett363157} by establishing a dependency on an upstream development project that is committed to supporting external users for its cloud-based services. However, the upstream project itself was being developed simultaneously, which increased the need for effective coordination between the two projects.

To implement this coordination, the collaboration established an Agile Scrum process for the development of the scientific software that tracked the ongoing Agile development of the upstream cloud-based service. We describe the coordination required to successfully manage these two parallel, but dependent, Agile processes, where the downstream process necessarily trailed the upstream one. We focus primarily on the downstream Agile process that this collaboration created and outline the effects this had on the ongoing, upstream Agile process. We also discuss the software engineering procedures we employed to ensure that the scientists could trust and depend on the in-progress upstream cloud-based service.

We review traditional and Agile software development procedures and provide an introduction to Agile abstractions in Section~\ref{sec:method}. We provide an overview the two groups in Section~\ref{sec:orgs}. We describe the creation of our collaboration and the risks involved in Section~\ref{sec:collab_dev}. Our approach to using Agile Scrum to resolve these risks is described in Section~\ref{sec:collab_struct}. We assess the collaboration's Agile Scrum framework and the outcomes for each organization in Section~\ref{sec:discussion}. We provide connections between this case study and other related works in Section~\ref{sec:related}. Section~\ref{sec:conclusion} summarizes the takeaways of this case study for the scientific computing community.

\section{Software Development Methodologies}\label{sec:method}
The integration of commercial software engineering best practices into scientific application development can be performed via a collaboration between scientists and professional software developers~\cite{storer2017bridging}. The goal of such a collaboration is for the scientists on the development team to use their domain expertise to explain software requirements to the software developers, allowing the two groups to utilize their collective knowledge to generate software requirements from the scientific use-case~\cite{segal2011challenges}. 

Such a partnership may introduce productivity risks for the professional developers, in addition to risks in the scientific development process. In this project, the professional development team had to maintain core development timelines and goals for its original stakeholder base, while at the same time developing new features and mechanisms to support the evolving scientific software development.

These issues can be aggravated by traditional software development frameworks such as the ``Waterfall'' approach, wherein the stakeholders and developers establish software requirements at the beginning of the project and then develop to meet those requirements~\cite{waterfall}. If the requirements change during the development period, the software is obsolete when delivered. To avoid this, Waterfall developers must correct the requirements and refactor the software during a release. If the requirements change rapidly enough or the refactoring time is too long, software delivery can be disrupted or delayed.

One approach to mitigate the issues of the Waterfall approach is to instead utilize an Agile Scrum software development framework~\cite{storer2017bridging,segal2011challenges}. The Agile software engineering process (described in Section~\ref{sec:agile}) is designed to be flexible to rapidly changing requirements. It is an iterative approach to software development, wherein week- to month-long development tasks are favored over large, long-term goals for the end product~\cite{Beck2013ManifestoFA,kane2006Agile}. Moreover, the goal of an Agile development effort is to have a releasable product at the end of each short development period, called a ``sprint''. Thus, as requirements change, the software is never more than a sprint's duration away from a partially-featured releasable state.

The Agile framework's emphasis on short-term goals might not seem reasonable for large-scale scientific projects that often have long-term science goals on the timescale of decades. The key observation in this context is that while the long-term science goals may remain relatively fixed, the requirements for the software that is developed to meet these science goals may change as scientific insights emerge or as personnel change. Agile software engineering encourages such incremental progress by organizing development into ``User Stories'', which are individual tasks created and completed as emergent software requirements are discovered~\cite{kane2006Agile,easterbrook2009engineering}. While some studies have combined the Agile and Waterfall methods~\cite{schuh2017agile} or utilize nested Agile sprints within a single team~\cite{sutherland2005future}, the use of a single sprint process is typical of Agile practices.

The Scrum approach is one method to adopt Agile practices into a project~\cite{segal2011challenges,kane2006Agile}. Scrum is a process framework that structures a project into a planning phase, design phase, and development phase to provide an incremental and emergent approach to Agile development~\cite{Schwaber95scrumdevelopment}. These elements are particularly relevant in the scientific computing communities given their tendency of rapidly-evolving software requirements~\cite{segal2011challenges,sletholt2011literature,segal2008models,segal2005software,crabtree2009empirical,lane2012theory}. Despite these potential advantages, Agile practices are not widely implemented~\cite{kelly2007software,storer2017bridging,umarji2009software} or incorrectly adapted~\cite{segal2011challenges} in the scientific computing community.

\subsection{Review of Agile Abstractions and Personnel Roles}\label{sec:agile}

An Agile development effort requires the participants to fulfill specific functional roles that manipulate a set of abstractions representing the software throughout its development life-cycle. We briefly summarize definitions of these abstractions and roles for those who may be unfamiliar with the terms. 

\paragraph*{Scrum} -- The overall organizational abstraction for an Agile project is the Scrum, typically comprising several components: \textit{sprint planning}, \textit{sprints}, and \textit{sprint retrospectives}. During a sprint planning meeting, the specific development tasks are discussed and decided upon. A sprint is a discrete period of software development during which the developers carry out the tasks identified during the sprint planning meeting. Finally, during a sprint retrospective meeting, the team reviews the previous sprint to identify uncompleted tasks and review team productivity.

A key element to this approach is that each sprint leads to the completion of the tasks identified during sprint planning even if the requirements change mid-sprint. That is, adapting the development in response to changing requirements occurs during sprint planning and sprint retrospectives. This is facilitated via daily check-in ``standing'' meetings involving all Scrum participants. During a check-in, each project member gives a short statement of what they worked on the day before, what they will work on in the current day, and any blockers where other tasks are limiting their progress.

\paragraph*{Abstractions} -- The principle abstractions in the sprint process are \textit{Epics}, \textit{User Stories}, \textit{tasks}, and the \textit{Backlog}. An Epic is an overarching functionality road-map for the software. It is expressed in terms of User Stories and the relationships between them. A User Story is a description of a specific functionality a user would like to have the software fulfill. Typically, a User Story consists of the identification of the user's role, the functionality, and the expected outcome. The collection of User Stories for the Scrum describe the expected user experience for the software. A task is a specific software development item that is necessary to meet the functionality desired by one or more User Stories. Each task is scoped so that it can be completed within the time allocated to a sprint and each developer takes responsibility for completing one or more tasks. The Backlog is the collection of tasks that have yet to be claimed by a developer in a given sprint. The Backlog is created during each sprint planning meeting and contains all tasks to be completed during that sprint. Thus, while each developer will work on their assigned tasks independently, the developers share the same Backlog. At the conclusion of a sprint, any uncompleted tasks remaining in the Backlog are analyzed during the sprint retrospective.

\paragraph*{Personnel Roles} -- A Scrum involves development personnel that fulfill one of three roles: \textit{Scrum Master}, \textit{Product Owner}, or \textit{Team Member}. Additionally, the project identifies \textit{Stakeholders} who represent those invested in the success of the software, but do not take an active role in the Scrum meetings. The Scrum Master runs the Scrum, chairing all meetings, ultimately creating the Backlog for each sprint, and managing the analysis of uncompleted tasks in a sprint retrospective. The Product Owner represents the interests of the Stakeholders and users. Finally, the Team Members comprise the developers who implement tasks during each sprint. 
\section{Organizations and Background}\label{sec:orgs}

The SuperNova Early Warning System (SNEWS) is an open, public alert system that is built to provide an early warning for core-collapse supernovae in the Milky Way galaxy~\cite{Antonioli:2004zb}. The first indication of a potential stellar explosion will be the arrival of a burst of neutrinos~\cite{Nakamura:2016kkl}. If several detectors report a potential supernova within minutes of each other, SNEWS will issue an alert to its subscribers, which include astronomical observatories, neutrino detectors, and amateur astronomers and citizen scientists.

There are seven neutrino experiments participating in the current SNEWS framework and about 20 neutrino detectors worldwide that are planned to participate in the new SNEWS~2.0 framework, which is the basis of this work. From these neutrino experiments, over 100 scientists from all over the world are part of the SNEWS collaboration. Each neutrino detector has different data formats that the SNEWS framework must be able to combine to identify neutrino burst coincidences. Additionally, these neutrino detectors will be active within SNEWS at different times. With these specific requirements, SNEWS is one of the few successful examples, to date, of cyberinfrastructure spanning major neutrino experiments~\cite{SNEWS:2020tbu}. In the current era of multi-messenger astrophysics (MMA), there are new opportunities for SNEWS to optimize its science reach for the next Galactic supernova. To achieve this goal, SNEWS depends on developing and sustaining software with a number of integrated components~\cite{SNEWS:2020tbu}. SNEWS received a three-year grant to complete this upgrade with the potential to extend the funding period. 

SNEWS represents one agent in an ever-growing community of MMA organizations that rely heavily on cyberinfrastructure to support their science requirements~\cite{Ivezi__2019,abbott2017multi,Graham_2019,smith2013astrophysical,Ageron_2012,huerta2019enabling,aartsen2017icecube,tns_website}. These requirements include prompt, reliable, and efficient alerting in addition to seamless support for different message formats and conventions~\cite{tns_website,barthelmy2000grb,Gabriel2006AstronomicalDA,rutledge1998astronomer}. The requirements also pertain to multiple detectors, institutes, and nations, and would be unrealistic for any individual group to address~\cite{Allen:2018yvz}.

\begin{figure}[t]
 \centering
  \includegraphics[width=0.99\columnwidth]{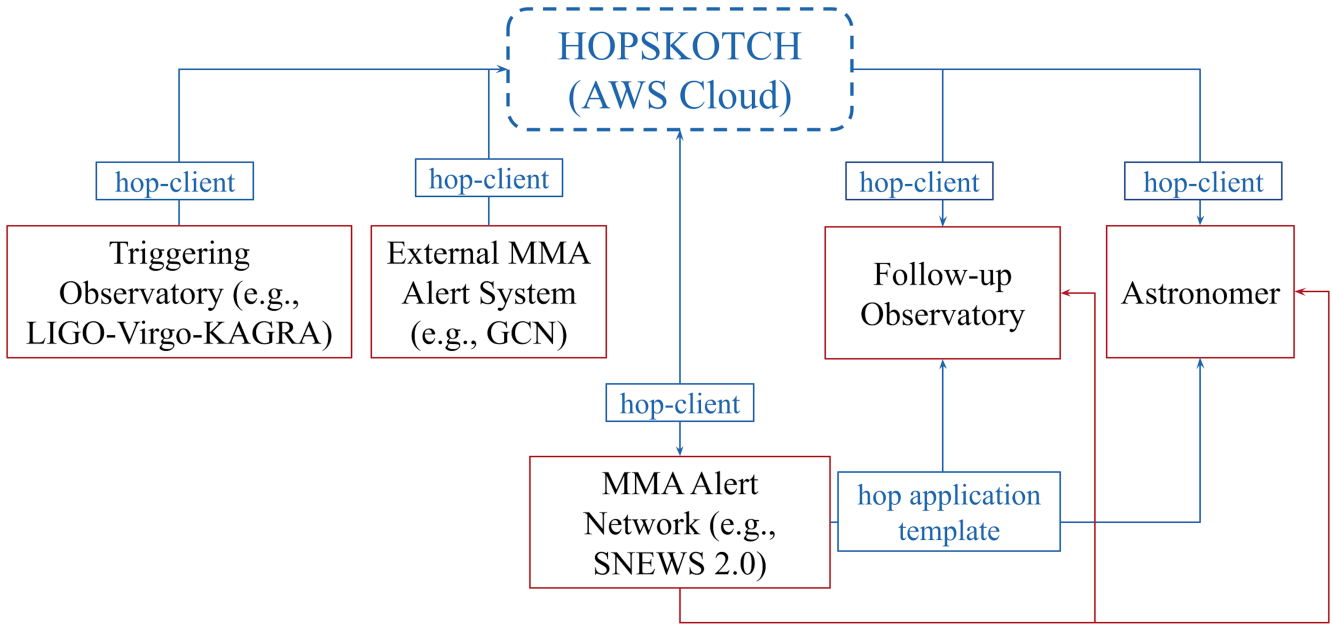}
  \cprotect \caption{Schematic of the services developed by SCiMMA (blue) and their integration into the MMA community (red). The \verb+hop-client+~\cite{patrick_godwin_2020_4033483} provides MMA observatories, systems, and users with authenticated publish/subscribe access to the \verb+HOPSKOTCH+~cloud~\cite{ScimmaAdmin,hopskotchSite} of Kafka topics. 
  }
  \label{fig:hopskotch}
\end{figure}

These challenges are not unique to the SNEWS effort. The MMA science community has developed various technologies, usually focused on a single project or experiment, that promote the sharing of results and collaboration among projects. However, MMA requires integration between individual systems and teams distributed across the world, highlighting a need for substantial cyberinfrastructure. For example, the landmark multi-messenger detection of the binary neutron star merger ``GW170717'' involved two gravitational wave detectors in the United States (the Laser Interferometer Gravitational-Wave Observatory detectors at Hanford and Livingston), one gravitational wave detector in Italy (Virgo), a gamma ray satellite (the Fermi Gamma-ray Space Telescope), and over six independent teams of astronomers~\cite{LIGOScientific:2017ync}. The coordination of such distributed teams and instruments requires a prompt, robust alerting system to facilitate communication between individual organizations. This requirement, and the general need for a common set of functionalities across otherwise separately-developed MMA systems, prompted the creation of the Scalable Cyberinfrastructure for Multi-Messenger Astrophysics (SCiMMA) project~\cite{Chang:2019edd}, a collaboration involving approximately 50 faculty, research scientists, and computing staff in physics, astrophysics, and computing from over 10 universities. SCiMMA has been addressing the cyberinfrastructure needs of multiple MMA organizations and science efforts by engaging with the user community~\cite{ScimmaGGroup,ScimmaWorkshop2020} and developing a suite of MMA services (Figure \ref{fig:hopskotch}) through a two-year community planning grant and three-year cyberinfrastructure grant. SCiMMA's development has focused initially on addressing the general need for the coordination of alerts and data between MMA organizations, such as the Vera C. Rubin Observatory's Legacy Survey of Space and Time (VRO LSST)~\cite{Ivezi__2019}, the advanced Laser Interferometer Gravitational-Wave Observatory (LIGO)~\cite{aasi2015advanced}, and the IceCube Neutrino Observatory~\cite{aartsen2017icecubeObs}. These services include the shared, openly available, multi-format data distribution software \verb+hop-client+~\cite{patrick_godwin_2020_4033483}, a customizable \verb+hop-client+ application template~\cite{ScimmaTemplate}, and a cloud-based system of data streams with extensive identity/access management controls known as \verb+HOPSKOTCH+~\cite{ScimmaAdmin,hopskotchSite}.

To develop these services, the SCiMMA collaboration employs an Agile Scrum framework staffed by 5-10 part-time scientific software developers, many of whom are already involved in the computing efforts of LIGO, VRO, and IceCube. The Agile team is led by two senior scientists and investigators from VRO and the North American Nanohertz Observatory for Gravitational Waves (NANOGrav)~\cite{nanograv}. SCiMMA is thus already connected to the MMA ecosystem, allowing it to deploy a set of successful practices and systems for the MMA community.~\cite{heroux2009barely,patrick_godwin_2020_4033483,ScimmaDocker}

\subsection{Collaboration Opportunity}

\begin{figure}[t]
 \centering
  \includegraphics[width=0.99\columnwidth]{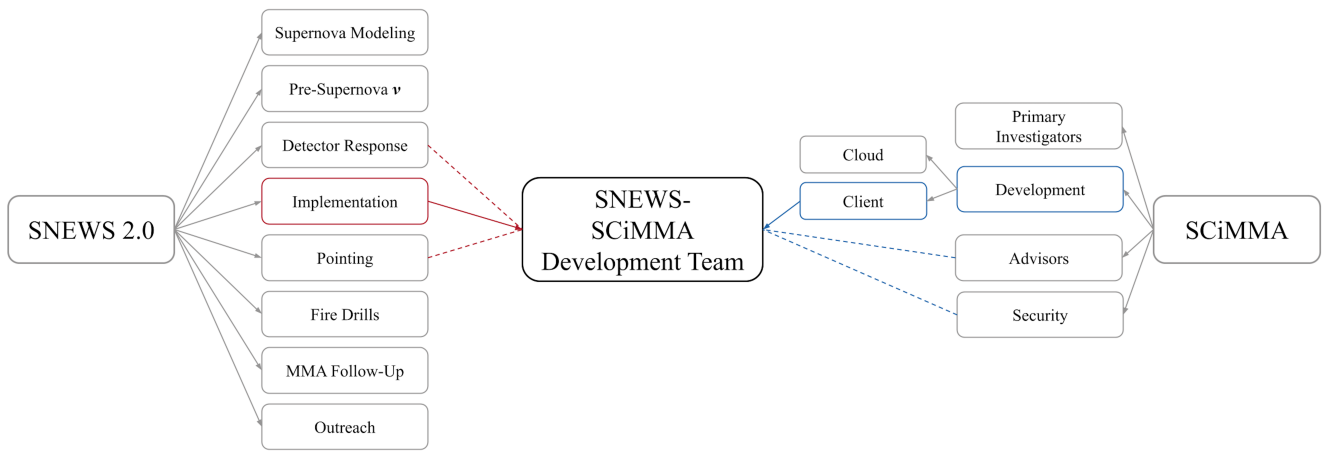}
  \caption{Collaboration ecosystems for SNEWS and SCiMMA. The SNEWS collaboration is broken into eight working groups. The SCiMMA collaboration is broken into four working groups. The SNEWS implementation group worked with the SCiMMA client-development group to form the development team described in this work. The dotted lines indicate teams which provided functional input to the development, but did not participate in the development itself.}
  \label{fig:ecosystem}
\end{figure}

SNEWS was preparing to upgrade its alert software framework from the current version of SNEWS to SNEWS~2.0. SNEWS~2.0 would require a scalable and reliable publish-subscribe system, potentially based on cloud-hosted web-services. The SNEWS team had already created the requirements for their first alert software as part of a Waterfall methodology and was exploring technology options for a purely in-house development effort~\cite{SNEWS:2020tbu}. Initially, SNEWS considered two options for their software development upgrade. SNEWS could either (1) build these systems from scratch, or (2) explore existing software. Each option would offer both advantages and challenges. Option 1 would allow SNEWS to maintain direct control over the development and ensure their needs were met, but it would impose the burdens of long-term development, support, and maintenance entirely on SNEWS developers. This could ultimately become problematic as SNEWS most likely would not have the resources and personnel to sustainably maintain the software long-term. Option 2 would avoid these burdens by offloading the core programming to an external software group, but it would reduce SNEWS's control in the process and potentially limit their representation and communication during development. 

At the same time that SNEWS was assessing the upgrade process, SCiMMA was in the midst of its first release cycle for new cyberinfrastructure systems, \verb+HOPSKOTCH+~and the \verb+hop-client+ (Figure~\ref{fig:hopskotch}), using an Agile Scrum methodology~\cite{hopskotchSite,patrick_godwin_2020_4033483}. These systems were developed as part of SCiMMA's goal of supporting the MMA community's need for sustainable, industry-standard software tools and cyberinfrastructure for alert data management and sharing, in order to support the use-cases of its current stakeholders from LIGO, VRO, and IceCube~\cite{Chang:2019edd}. 

The connection between SNEWS and SCiMMA began when a principle investigator from SNEWS met with leadership from the SCiMMA team at an in-person National Science Foundation ``Harnessing the Data Revolution'' grant meeting~\cite{hdr_nsf}. During the discussions, it became clear that \verb+HOPSKOTCH+ either provided or would provide the functionality required by the SNEWS~2.0 system, and that SNEWS could serve as SCiMMA's first true user group. While none of the developers from SNEWS and SCiMMA had met, the members shared a common language of scientific software development and the principle investigator discussions indicated that cooperation could lead to potentially-mutual benefits. Given this clear overlap of interests, the two groups proposed an informal, remote cooperation between SNEWS and SCiMMA.

Initially, the cooperation began experimentally without direct collaboration between the groups. To test the hypothesis that \verb+HOPSKOTCH+ would meet SNEWS's requirements, the SNEWS implementation team carried out their own development using publicly-available code from SCiMMA~\cite{patrick_godwin_2020_4033483,ScimmaTemplate} without directly engaging the SCiMMA client development team. This allowed SNEWS to explore the potential benefits of SCiMMA's systems without significant financial investment, as this required only partial effort from a handful of individuals without the need for a separate grant or funding source. To become more familiar with the code, the SNEWS team engaged with one of SCiMMA's virtual community workshops to further assess the value of a collaboration~\cite{ScimmaWorkshop2020}. This phase was a traditional example of a user-group adopting a piece of open-source software, wherein developers modify another group's existing codebase to fit their own needs without coordination or direct interaction between the groups.

Once SNEWS developers began integrating \verb+HOPSKOTCH+ into their early prototype of SNEWS~2.0, they found that they required more extensive features that were not available in the current version of SCiMMA's systems. SNEWS already knew exactly what features they needed in SCiMMA's software, but did not have the developer effort necessary to implement them. SCiMMA had ample professional software development effort to provide for this need, but did not have an understanding of SNEWS's specific scientific requirements and software needs. Given this split of domain knowledge and software development talent between the organizations, the traditional open-source development strategy of making feature requests or independently customizing the software would not meet SNEWS's needs. Therefore, SNEWS principle investigators contacted SCiMMA principle investigators to discuss the possibility of a focused collaboration to develop the desired features together. This prompted the creation of a partnership that evolved into a dedicated bilateral collaboration involving members from both the SNEWS implementation team and the SCiMMA client-development team (Figure~\ref{fig:ecosystem}). This combined effort allowed a collaboration to form via piece-wise funding contributed by both SNEWS and SCiMMA, without the need for a formal joint funded proposal between the organizations. This direct engagement ultimately catalyzed development for both organizations, but both teams were concerned with the inherent risks due to the collaboration's novel structure and the fact that each organization's development was at an early, prototyping stage. The main concerns were:

\begin{itemize} 
\item Can SNEWS rely on pre-alpha, externally-supported systems to sustainably accomplish the scientific goals for SNEWS~2.0? Would the scientific requirements be communicated and addressed adequately on a short time-frame in a collaboration? 

\item Can SCiMMA temporarily prioritize a single user-group without derailing the ongoing development needed to support the broader MMA community? Would early user-engagement be feasible and beneficial in the long-term? 
\end{itemize}

These risks could not be confronted until the SNEWS developers first ensured that SCiMMA's work would be useful in their own efforts. Since SCiMMA had made its codebase openly available and promoted its development efforts via community workshops, SNEWS could assess an active collaboration with SCiMMA prior to committing to such an engagement. We believe that this initial trust-establishing step was key to the ultimate success of the effort. Without it, much more of the collaboration itself would have focused on trust building and maintenance rather than on productivity and risk management.

\section{Collaboration Apprehension} \label{sec:collab_dev}

\begin{figure}[t]
 \centering
  \includegraphics[width=0.99\columnwidth]{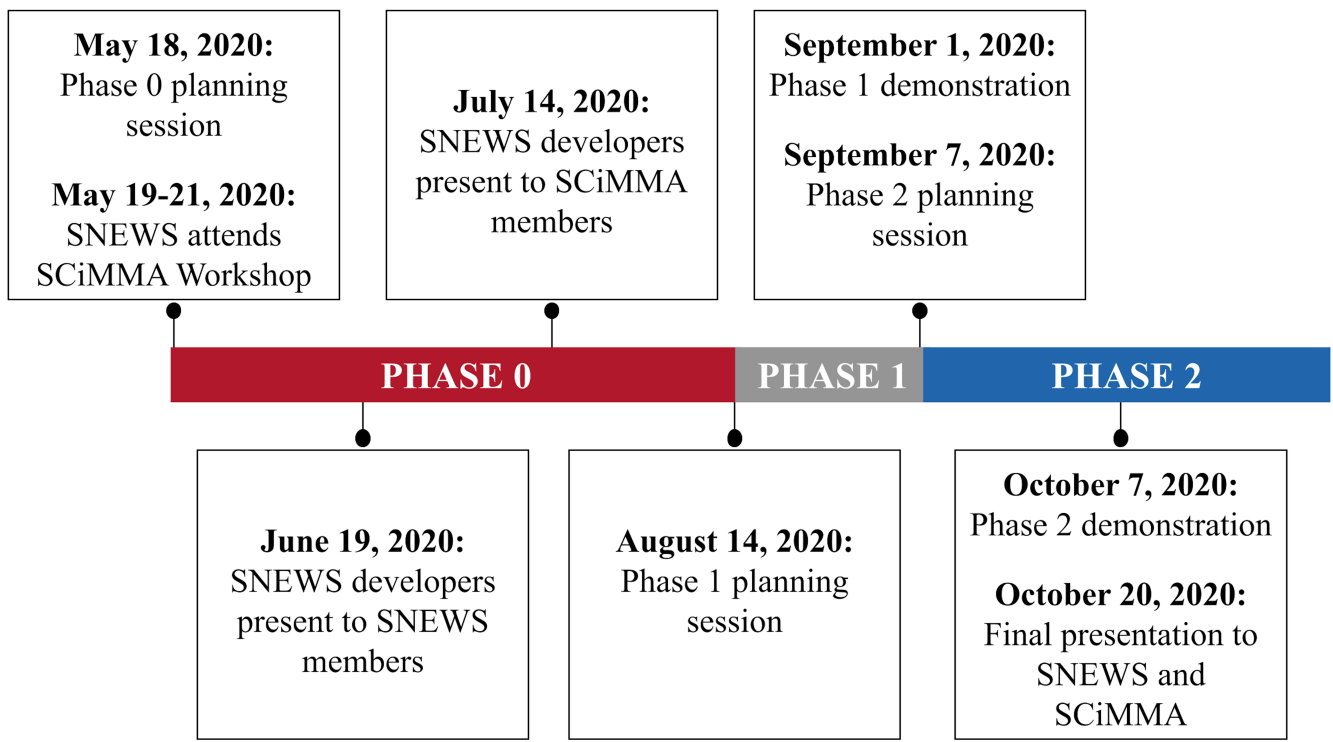}
  \caption{Timeline of events that led to the development of the SNEWS-SCiMMA collaboration and the SNEWS~2.0 prototype. Indicated on the timeline are the three phases of the collaboration. The conclusion of each phase marked an evaluation period for each organization, which provided a decision point on whether to continue the collaboration.}
  \label{fig:timeline}
\end{figure}

While the collaboration presented a unique opportunity to accelerate development, it brought risks to both teams. The concerns of each group provided an outline for the nature of the collaboration, which evolved through several stages of bilateral development effort and progress evaluations. To address these concerns (Section \ref{sec:collab_struct}), the collaboration adopted its own Agile Scrum framework in parallel with SCiMMA's own Agile Scrum process. This new weekly sprint trailed SCiMMA's ongoing bi-weekly sprint. This promoted communication and drove software development in response to changes introduced by both the SNEWS requirements and the original SCiMMA stakeholders. 

\subsection{Concerning the reliance on external software} 

Upon learning about SCiMMA and \verb+HOPSKOTCH+, SNEWS considered using these services for their own development prototyping. This would allow SNEWS to actuate its initial development using the already-existing systems and resources provided by SCiMMA, which would be supported and maintained by professional software developers long-term. However, the main concern of the SNEWS team was that the development priorities of SCiMMA may not align with SNEWS's feature requests and bug reports, which could cause a potential latency or disconnect between SNEWS's requirements and SCiMMA's implementation. Moreover, the software could become unmaintainable if the SNEWS team wasn't engaged in the development process. Nonetheless, SCiMMA's available features and end goal appeared to fit all of the initially identified needs, so SNEWS decided to proceed.

SNEWS carried out the initial development which began in Phase 0 (Figure~\ref{fig:timeline}). During this stage, SNEWS developers attended the SCiMMA 2020 Virtual Workshop software demonstration session~\cite{ScimmaWorkshop2020}. This allowed the SNEWS developers to utilize the SCiMMA \verb+hop-client+ application template, a customizable interface to \verb+HOPSKOTCH+, as the basis to prototype SNEWS~2.0 using SNEWS's previous software requirements. They extended the template with additional SNEWS-specific functionality, maintained the codebase in a private GitHub repository, and ran local tests in Python. During development, SCiMMA added the SNEWS developers to the SCiMMA Slack~\cite{slack} channel to provide user support and debugging as SNEWS developed the prototype, though the development itself was driven entirely by SNEWS. This phase resulted in a successful re-deployment of SNEWS's basic functionality using SCiMMA's \verb+hop-client+.

Once SNEWS developed their prototype to the extent possible with \verb+hop-client+, SNEWS presented a status update at SCiMMA's July 2020 Public Teleconference ~\cite{ScimmaTelecon_July2020}. This presentation highlighted SNEWS's current implementation, usage requirements, and desired features as realized during Phase 0 (Figure~\ref{fig:timeline}). Some requested features were not currently in SCiMMA's software, such as a database for message storage or a customizable message format. The SNEWS team considered three options: develop the missing requirements themselves, wait until SCiMMA developed these requirements in their own timeline, or seek a collaboration with SCiMMA to develop the requirements together. After this status update, it was not clear which path forward SNEWS should take.

\subsection{Concerning available development effort}

The SNEWS status update prompted a period of internal assessment for SCiMMA to evaluate the feasibility and utility of collaborating with SNEWS. The potential benefits of a collaboration were promising, as SNEWS would serve as SCiMMA's first development partner and provide a practical use-case for its cyberinfrastructure. This could accelerate the development and demonstrate wider community impact for SCiMMA~\cite{Chang:2019edd}. SCiMMA therefore proposed a short period of focused development with SNEWS after the SNEWS status update near the end of Phase 0. However, this collaboration presented a risk to SCiMMA's core goals due to its limited funding: SCiMMA's cyberinfrastructure grant had to support \verb+HOPSKOTCH+ cloud costs in addition to the developers. To mitigate this risk and ensure that development would not be derailed, the SCiMMA team formally integrated the collaboration with SNEWS into its own Agile sprint structure. The team created an Epic and User Stories specifically for SNEWS development support, and assigned this project to two developers. These SCiMMA developers remained active in core SCiMMA tasks and assessed the SNEWS codebase, but did not commit any time towards SNEWS software development during the remainder of Phase 0. After SNEWS's presentation to SCiMMA and the SCiMMA developers' internal sprint presentation, it became clear that a collaboration would support SNEWS's development goals while providing an informative use-case for SCiMMA's own development. While there were still concerns about allocating effort away from core development, SCiMMA proposed a short period of focused collaboration with SNEWS (Phase 1 in Figure~\ref{fig:timeline}) given the many synergies between the projects. Both groups then decided to actively work together to further the development of the SNEWS~2.0 prototype.

\subsection{Concerning communication during a collaboration}

The assessments during Phase 0 confirmed both that \verb+HOPSKOTCH+~was a valuable technology for supporting SNEWS~2.0 and that SCiMMA was willing to work with SNEWS to meet their requirements and timeline, suggesting that a collaboration would be mutually beneficial. However, given the concerns of each organization, the structure of the collaboration would need to meet several constraints. Foremost, the workload would need to be shared between organizations. This would ensure that SNEWS would have the understanding required to sustain the codebase after SCiMMA's involvement, allow the SCiMMA developers to understand the existing SNEWS codebase on a short timescale, and avoid software duplication. This collaboration would also change SCiMMA's previous community engagement model, wherein MMA organizations served only as end-users rather than collaborative developers. This would increase the importance of reliable communication between SNEWS and SCiMMA to identify and evaluate ideas emerging from the collaborative development. The emergent nature of these constraints, along with the short timescale, suggested that an Agile Scrum framework would benefit the collaboration. However, the addition of a second Agile Scrum process to SCiMMA's effort would require careful implementation to avoid complicating the separate, but dependent, development efforts. Moreover, trust would have to be built in the Agile process itself since none of the SNEWS team had utilized Agile methods for software development nor heard of its success in scientific computing.

\section{Using Agile Scrum to Mitigate Apprehension} \label{sec:collab_struct}

The collaboration was structured to address the concerns of SNEWS and SCiMMA to ensure that the combined effort would benefit each individual group. While the shared funding commitment from SNEWS and SCiMMA reduced some of the up-front risk of collaborating, mutual trust was still required in order to address the concerns of each organization. The success of one group was dependent on the success of the other, so each team needed to consider the other's requirements in addition to their own. This promoted mutual interest between the two projects and provided a foundation of investment and trust despite the lack of in-person meetings, pre-existing social connections, or formal joint funding proposals. The pre-collaboration planning (Phase 0) and the initial two-week period of focused collaborative development (Phase 1) provided informative requirements for the collaboration structure, which evolved into a more formal Agile framework in the four-week Phase 2. Each phase proceeded only after a consensus between organizations was reached, so that the conclusion of each phase allowed each organization to build trust and evaluate whether to continue the collaboration. This approach ensured that the collaboration was addressing the evolving needs and perspectives of each organization.

This collaboration was composed of two members of SCiMMA and four members of SNEWS, from the working groups pictured in Figure~\ref{fig:ecosystem}. In this collaboration, the Scrum Master was a part-time SCiMMA software developer who primarily worked on gravitational-wave computing cyberinfrastructure. There were two SNEWS Product Owners, one primary investigator and one graduate student from particle astrophysics domains. The Scrum Master and Product Owners identified the specific User Stories that were to be tasked during each sprint. The entire team included the Scrum Master and Product Owners, a second software developer from SCiMMA, an undergraduate student developer from SNEWS, and an academic IT specialist from SNEWS. This composed the core Scrum team, which coordinated sprints via sprint planning sessions, sprint retrospective meetings, and daily check-ins. The Stakeholders included six primary investigators in total, three from SCiMMA and three from SNEWS, who served to evaluate the phases and decide if a new phase should be generated. Due to the geographically-distributed, remote nature of the collaboration, all Scrum and Stakeholder meetings were conducted via Zoom~\cite{zoom}, check-ins were facilitated via the online real-time messaging system Slack~\cite{slack}, and User Story management was conducted via a GitHub Project Board~(Figure \ref{fig:board})~\cite{board}.

\subsection{Addressing simultaneous development efforts}

As the Phase 1 collaboration began, SCiMMA created an Epic in its Agile process for a collaboration with SNEWS that would run in parallel to SCiMMA's main development efforts and dedicated two SCiMMA developers on this Epic. Sprint deliverables from this SCiMMA Epic triggered the addition of User Stories to the SNEWS Backlog. SCiMMA integrated the SNEWS collaboration progress into the regular SCiMMA Agile Scrum status updates during daily check-ins and sprint retrospectives, ensuring that the SNEWS collaboration would be synchronized with SCiMMA core development. The regular status updates provided an active dialogue throughout Phase 1, allowing SCiMMA to remain in-sync with emergent feedback and ready to adjust its development plans in response. This dialogue involved multiple new User Stories generated from bugs and feature requests during Phase 1. In addition, SCiMMA noted that many of these User Stories were of general MMA interest and decided to transfer some of these stories from the SNEWS Backlog to their own Backlog. This synergy suggested that dedicating more effort for a focused collaboration could continue benefiting core SCiMMA development, even if it reduced total effort available in the short-term. Since these potential benefits were within SCiMMA's primary scope, the developers could transition to collaborative development in support of SCiMMA's preexisting Agile process with little concern about formalizing a commitment of time and funding away from core efforts. Moreover, core development could still proceed even if several developers were focused exclusively on an external collaboration, as the rest of the team could continue working on components that were independent of the SNEWS client development, such as the \verb+HOPSKOTCH+ cloud servers.

\subsection{Addressing communication \& reliance on external software}

To implement Agile methods for Phase 1, the Scrum Master coordinated meetings and the creation of User Stories based on the Product Owners' scientific requirements. Developers from each team actively engaged during Phase 1, communicating regularly via Slack check-ins, Zoom meetings, and by using a shared Github project board. The two teams conducted one-week sprints, consisting of a review meeting each week to informally check progress and create User Story task cards on the project board. These meetings continued in-parallel with SCiMMA's ongoing bi-weekly sprint meetings, with additional virtual impromptu pair-programming sessions for code debugging and technical questions. The Scrum Master regularly consulted with the SCiMMA Scrum Master and Product Owners during their ongoing sprint process. The SNEWS Product Owners were responsible for ensuring that the SNEWS~2.0 prototype was within specifications, communicating the needs of SNEWS to SCiMMA during sprints, and reporting the collaboration progress during SNEWS's internal meetings. The other developers worked on improving the prototype and deploying it on test systems. This framework ensured bilateral effort for the collaboration and provided regular communication between the organizations. It gave SNEWS an active role in the development, allowing for feature requests and bugs to be relayed through the SCiMMA developers to the ongoing SCiMMA sprint. This also allowed for SNEWS to become familiar with the software, reducing the risk of unsustainable software after the collaboration. The Agile Scrum framework also supported timely development by allowing large goals to be translated into smaller tasks, promoting iterative development.

\subsection{Addressing timeline constraints}

Phase 1 concluded with a demonstration of the software to the SNEWS and SCiMMA stakeholders. After reviewing the collaboration progress, it was clear that there was mutual benefit and that the initial concerns appeared to be ameliorated in a manageable way. Both organizations agreed to a second, four-week period of focused development (Phase 2 of Figure~\ref{fig:timeline}). However, at the end of Phase 1, there were clear timeline constraints: SNEWS was seeking to begin the transition to production operation of the software and SCiMMA expected to focus more effort on the internal development tasks that had resulted from the collaboration. These timelines were constrained by the funding available for each project independently. These new constraints prompted several changes to the collaboration structure.

\begin{figure}[t]
  \centering
  \includegraphics[width=0.90\textwidth]{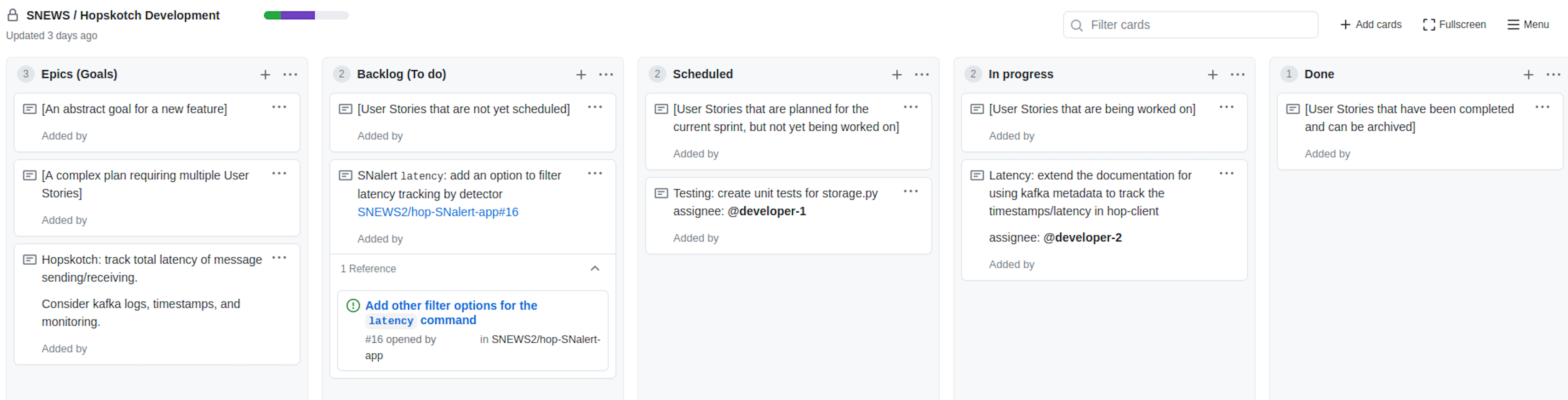}
  \caption{An example of the GitHub Project Board used for the SNEWS-SCiMMA development collaboration. At the beginning of Phase 1 and 2, developers would brainstorm Epics in the first column based on feedback from the two organizations. During each sprint, the Scrum Master would work with developers to generate new User Stories from the Epics. These User Stories would remain in the \textit{Backlog} column until they were assigned to a specific developer and later moved to the \textit{Scheduled}, \textit{In-progress}, or \textit{Done} columns over the course of the sprint.}
  \label{fig:board}
\end{figure}

Phase 2 continued to use the GitHub Project Board and daily check-ins on Slack as in Phase 1, but the sprints were restructured to better reflect timeline constraints. Four one-week sprints were set up, with a weekly sprint meeting consisting of review and planning sections, in addition to the ongoing daily check-ins. In the review section, the developers gave short, round-table updates on their progress and blockers from the previous week based on the cards they were assigned. In the planning section, the Scrum Master evaluated the status of the \textit{Scheduled} and \textit{In-progress} task cards in the project board~(Figure~\ref{fig:board}), marking them as \textit{Done} or \textit{In-progress}. The developers then created new Backlog cards by discussing the existing Epics and self-assigned them to work on in the upcoming week. This revised format matched the Agile Scrum development model, as the short-term iterative development progress could be checked against the longer-term Epics during each week.

While the one-week sprint cadence differed from SCiMMA's two-week sprint schedule, this allowed the SNEWS effort to stage its planning and evaluation meetings both between SCiMMA sprint meetings and also immediately after them. In this way, the SNEWS effort was able to respond to requirement changes introduced by the core SCiMMA effort. If the Agile timelines had been the same, there was a risk that the SNEWS effort would respond during a two week sprint only to find that the core effort had already introduced specification changes, and thus SNEWS would never be able to remain in-sync with the SCiMMA development. This allowed SNEWS to keep a separate timeline for its own deliverables without depending on the timeline for core SCiMMA development.

The second purpose allowed SNEWS to pick a specific release date set for the end of Phase 2. SNEWS's goal for Phase 2 was a product that could be effectively released to SNEWS for further enhancement and operational deployment. After this release, SNEWS would contribute to SCiMMA as a community member through existing interaction mechanisms such as workshops, but not through direct collaboration. Thus the developers on both teams decided on multiple week-long sprints to achieve the end goal for the software. As with Phase 1, Phase 2 concluded with a demonstration of the software to the stakeholders, which was recorded as an endpoint marking the conclusion of the dedicated software development period~\cite{ScimmaTelecon_October2020}.

\section{Discussion} \label{sec:discussion}

\begin{figure}[t]
  \centering
  \includegraphics[width=0.90\columnwidth]{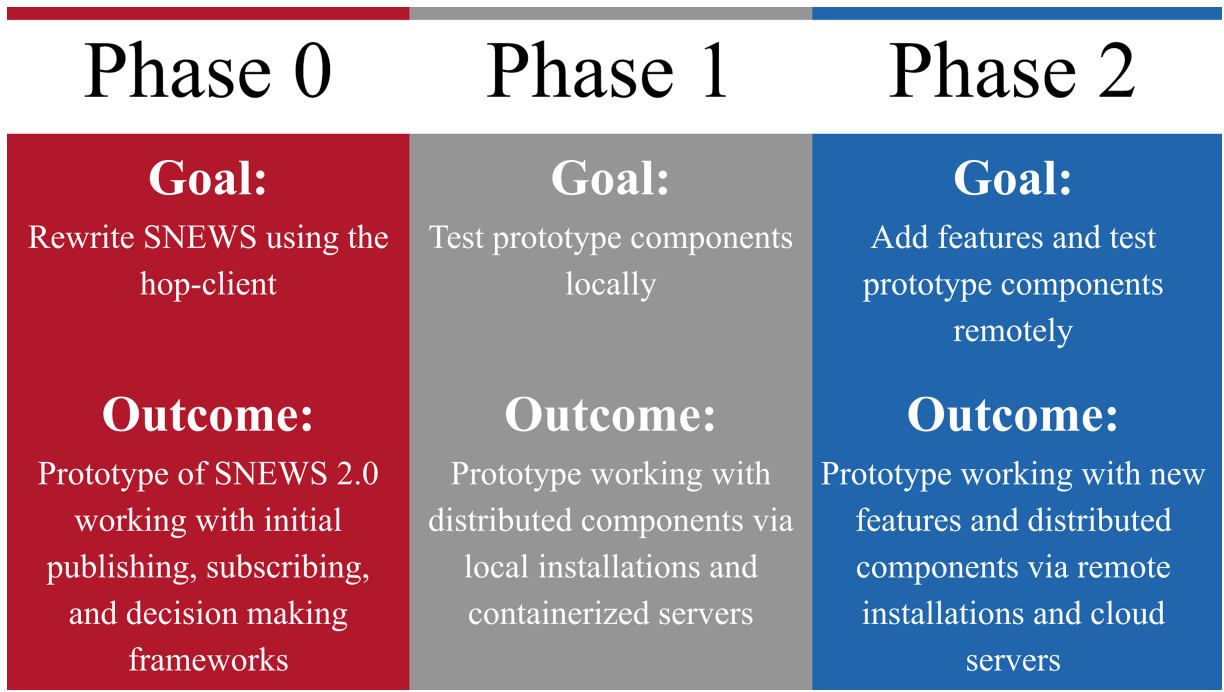}
  \caption{Illustration of the goals and outcomes for each phase of the SNEWS-SCiMMA collaboration. Each phase proceeded after a progress review and consensus within each organization.}
  \label{fig:phases}
\end{figure}

Each phase of the collaboration was structured around technical demonstrations to iteratively evaluate progress, assess current goals, and plan out goals for the next phase. These demonstrations allowed for trust-building and re-evaluation of personnel commitments and funding for each organization. Phases 1 and 2 utilized weekly meetings coordinated via a GitHub Project Board (Figure~\ref{fig:board}) for task and project management. The evolution of the codebase and development metrics throughout the collaboration is outlined in Figures~\ref{fig:codedelta}~and~\ref{fig:metrics}, allowing for the success of the collaboration to be tracked throughout the phases.

\subsection{Measurements of Success}
The goal of Phase 0 was to rewrite the current SNEWS publish-subscribe system using the \verb+hop-client+~(Figure~\ref{fig:phases}). Though an Agile framework was not used during this phase, Phase 0 served a similar role to the Agile Inception Phase: planning and deciding on software requirements was the primary goal. Phase 0 was marked successful by the demonstration of a working prototype that included sample publishing and subscribing of messages with the SNEWS decision making framework~\cite{ScimmaTelecon_July2020}. This established the system's baseline prior to any collaborative development and gave SNEWS the confidence to rely on external software. 

Phase 1 was then planned around a goal to develop a prototype of the system running locally~(Figure~\ref{fig:phases}). To capture SNEWS's large goals from Phase 0, the developers created five broad Epics. These were translated into 10 User Stories over the course of Phase 1. Some of these Stories were relatively broad and took more than one sprint or developer to complete since the teams were learning how best to collaborate and manage the project. Nonetheless, eight of the Stories were completed by merging five pull-requests and resolving four issues during Phase 1, as illustrated in Figures~\ref{fig:codedelta}~and~\ref{fig:metrics}. The teams assessed the progress of this phase via a demonstration of a prototype that used local containerized services and a local installation of the SNEWS~2.0 app. The prototype at this stage replicated most of the current SNEWS functionality. Both teams approved of the collaboration's progress: SNEWS had a new implementation of their server and SCiMMA improved the \verb+hop-client+ features beyond their initial scope. This phase and its success enhanced the trust between the two teams, encouraging the teams to agree to a second round of collaboration, Phase 2.

Phase 2 built on the Phase 1 goal of a local prototype by extending the prototype using distributed systems, with cloud servers to demonstrate robustness of remote but integrated components~(Figure~\ref{fig:phases}). The developers reviewed the remaining Epics and User Stories from Phase 1, and determined that two Epics were no longer in-scope with the teams' goals after the Phase 1 demonstration. The remaining three Epics and two User Stories were determined to still be relevant and were carried over into Phase 2. This phase began with six new Epics based on the assessment of the Phase 1 demonstration, focusing on measuring the prototype's performance and improving client authentication. This phase allowed SCiMMA to test a service that used their cloud servers and SNEWS was able to create a fully-functioning prototype. The Epics were translated into 29 User Stories over the course of Phase 2. Of these, 15 were completed by merging eight pull-requests and resolving three issues during Phase 2. Both organizations reconvened for a demonstration of the remote prototype, which met the initial goal and concluded Phase 2~\cite{ScimmaTelecon_October2020}. After the demonstration, seven of the nine Epics were completed. The remaining two were either ill-defined or out-of-scope for the collaboration. Fourteen of the 29 User Stories were incomplete: five \textit{Backlog}, six \textit{Scheduled}, and three \textit{In-Progress}. The developers continued working on four of these User Stories independently after the collaboration; the remaining 10 User Stories were considered out-of-scope for the collaboration period since the desired Epics had already been met.

The three phases resulted in a variety of development metrics that progressively evolved during the collaboration. In particular, while the SNEWS developers were able to create a prototype independent of SCiMMA by making moderate additions to the codebase (Phase 0 of Figure~\ref{fig:codedelta}), the collaboration with SCiMMA (Phases 1 and 2 of Figure~\ref{fig:codedelta}) demonstrably accelerated development. In particular, Phase 1 resulted in roughly half the amount of the codebase changes made during Phase 0 in a quarter of the time (two weeks, compared to eight); Phase 2 then led to even more changes than either Phase 0 or Phase 1. While this may in part be due to the simple increase of the number of developers, Figures~\ref{fig:codedelta} and~\ref{fig:metrics} suggest that the Agile practices and cooperation could offer added benefits.

The progression of the phases reflects the evolving Agile coordination in this collaboration. Phase 1 was fruitful, but remained a short-term trust-building exercise. The demonstrated success of Phase 1 encouraged both teams to invest more resources, effort, and time into the collaboration and provided a basis for Agile methods that Phase 2 continued to develop. This is shown in the evolution of Agile metrics across each phase in Figure ~\ref{fig:metrics}. The results of Agile practices remained steady during Phase 1 and 2, wherein the four-week Phase 2 yielded roughly twice the number of Epics and User Stories compared to the two-week Phase 1. This demonstrates that the short-term success of Phase 1 could be extrapolated to a longer engagement; note, however, that Phase 2 involved more organized Agile practices compared to Phase 1, which may suggest that the less-formal implementation of Agile in Phase 1 may not have been adequate for a longer phase.

\subsection{Future Directions}
\begin{figure}[t]
\begin{minipage}[t]{3.4in}
    \includegraphics[width=\columnwidth] {fig_codedelta_dpi.png}
    \caption{Lines of code changed during each phase of this case study. Positive values (blue) are the number of lines of code added to the SNEWS~2.0 codebase. Negative values (red) are the number of lines removed from the codebase.}
    \label{fig:codedelta}
\end{minipage}
\hfill
\begin{minipage}[t]{3.4in}
    \includegraphics[width=\columnwidth]{fig_metrics_dpi.png}
    \caption{Figure displaying the software development and Agile framework metrics during each phase. We show the number of commits (dark red), issues resolved (orange) and merged pull requests (light red). Additionally shown are the number of epics created (light blue), user stories created (blue), and number of user stories completed (dark blue).}
    \label{fig:metrics}
\end{minipage}
\end{figure}

\begin{figure}[t]
 \centering
  \includegraphics[width=0.99\columnwidth]{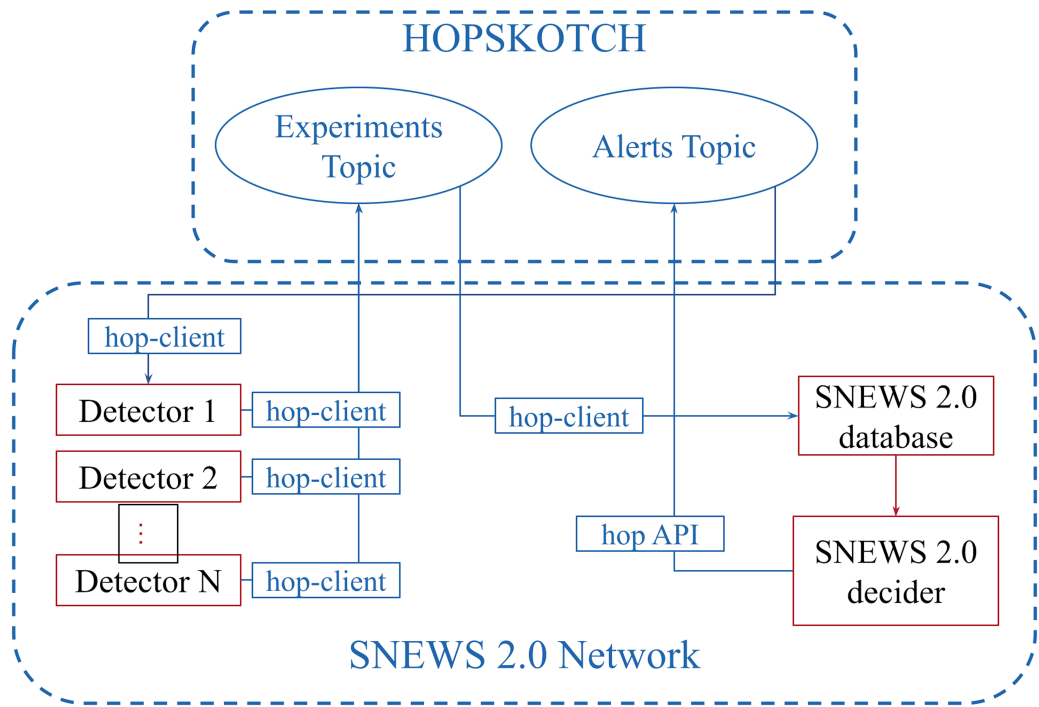}
  \caption{Schematic of the SNEWS~2.0 prototype application. The application was developed as an extension of the hop application template to contain a database and coincidence-checking system. }
  \label{fig:snews2}
\end{figure}

After the demonstration and evaluation of Phase 2, the collaboration had successfully produced a prototype of SNEWS~2.0 using SCiMMA's \verb+hop-client+ publish-subscribe system~\cite{patrick_godwin_2021_4437579}. A schematic of the prototype is shown in Figure~\ref{fig:snews2}. The prototype established directions for future development for each group, allowing the organizations to remain in contact even after the phases of focused collaboration had concluded. It has provided direction for SCiMMA's authentication and permission management infrastructure, and has helped SNEWS begin prototyping their detector network using the SCiMMA \verb+hop-client+. The next steps after the development of the prototype include having SNEWS members from outside the development team test the software using tutorials provided by the SCiMMA developers to ensure that the software is user-friendly. After this testing, SNEWS plans to integrate all of its neutrino experiments into the SNEWS~2.0 network. This will happen during a dedicated workshop wherein SNEWS will implement Agile methods, similar to those utilized during Phase 2 to drive the development. Specifically, the workshop will take place over a four-week sprint and will include daily check-in meetings to discuss progress. Additionally, Epics and User Stories will be defined to keep the integration plans on track.

The collaboration promoted the general development efforts for each group. SNEWS utilized SCiMMA's systems and software engineering expertise to quickly develop their prototype, allowing SNEWS to avoid the difficulties of scientific computing and software development. The collaboration also unexpectedly accelerated the pace of SCiMMA's core development since SNEWS' use-case contained features that were of broader MMA interest. Additionally, it provided an exemplar for future engagements: SCiMMA plans to integrate Agile in future collaborations within the MMA community.

\subsection{Lessons Learned}

The success of this collaboration relied heavily on the Agile Scrum framework. The implementation of Agile practices in our work provided several key elements that ameliorated the difficulties of scientific software development as well as the concerns that emerged from each organization, which suggests that some of the Agile practices specific to our collaboration may be broadly relevant to the scientific software community. 

First, the daily Slack check-ins and weekly sprint meetings facilitated constant communication between otherwise independent developers. This ensured that the emergent concerns and evolving scientific requirements of SNEWS and SCiMMA were brought up and addressed. The team found that the week-long sprints and User Story planning helped identify achievable tasks within broader goals of the collaboration. In particular, the sprint and phase retrospective meetings were crucial to the success of this collaboration by allowing the team to frequently evaluate its progress and the resulting software to ensure that the needs of both SNEWS and SCiMMA were being addressed. The progress reports at the ongoing SCiMMA sprint within phases and the collaboration-wide leadership meetings between phases ensured that the separate sprints remained on track. Having persistent communication channels was also essential since the development team was geographically distributed and entirely remote for the duration of the collaboration.

Second, implementing an Agile framework accelerated the trust establishment. Trust in the Agile process itself was also essential to the collaboration, since while the SCiMMA developers were already familiar with Agile methods, the Scrum Master did not have prior experience with running a sprint and the SNEWS developers were entirely unfamiliar with Agile. The sprint meetings and required daily communication ensured that each member had regular correspondence with the others, which allowed the otherwise unacquainted teams to quickly meet and trust the other remotely. This demonstrated the advantages of Agile to the collaboration, enhancing trust in the sprint process itself. Additionally, the trust establishment from the Agile framework built upon the shared scientific backgrounds that the two teams had: nearly all of the developers were already involved in scientific computing and/or multi-messenger astrophysics. The Agile sprint process capitalized on this mutual scientific language to further enhance trust and communication regarding the scientific software development effort. In particular, the SNEWS Product Owners were able to rapidly convey the technical alerting requirements of the application to the SCiMMA developers since the developers were already familiar with alert software in another area of astrophysics. Likewise, the SCiMMA developers brought technical software engineering expertise that allowed them to work with SNEWS computer scientists and system administrators. In particular, they worked with the computer science student developer to enhance the packaging, documentation deployment, and unit testing systems for the software, as well as the academic IT specialist to deploy the software on local hardware.

Third, the coordination of two Agile processes was novel to both organizations. We believe that adequate trust evolved nonetheless due to two reasons. First, the Agile process continuously generated measurable artifacts in the form of User Stories. The Product Owners and various stakeholders auditing the process could see, in real time, the progress and thus were willing to increase their trust. The second reason is that the coordination increased communication between the different organizations, which was particularly important since the development teams were involved in separate sprints and Agile alone is intended to orchestrate a single sprint team. By coordinating the downstream process via User Stories generated from the upstream process and holding cross-sprint check-ins, we created a method to communicate efficiently between the development teams.

\section{Related Work}\label{sec:related}
Our work serves as another case study within the rich ecosystem of scientific software development~\cite{Hannay2009,segal2005software}, Agile methods utilized by scientists~\cite{sletholt2011literature,easterbrook2009engineering}, and non-traditional Agile processes for software development~\cite{brenner2015scaled,schuh2017agile,sutherland2005future}.

Our work shares much with the work described in Ref.~\cite{segal2005software} in that both describe experiential case studies involving interactions between scientists and software engineers. Our work differs in that the teams involved do not belong to a single organization nor are the teams focused on the completion of a single goal. Further, the author of Ref.~\cite{segal2005software} cites difficulty in creating effective specifications as a key impediment to a successful collaboration, while for our project, the adoption of Agile alleviated this difficulty. However, the collaborative element of our Agile process introduced additional risks to the success of our work that were not covered in these case studies.

In Ref.~\cite{Hannay2009}, the authors conduct and report on a survey of scientific development teams and organizations to determine how these teams build software. They report features of scientific development described in their survey that our collaboration also embodies, such as the ``trust'' in community-maintained software when the community is large, such as SCiMMA's use of public Kafka libraries~\cite{ConfluentPython}. Our work is largely complementary, however, as we adopted a specific approach without a controlled comparison with other alternative approaches. As such, our work would constitute a single, successful data point within this previous survey. 

In Ref.~\cite{easterbrook2009engineering}, the authors survey climate scientists to determine their common software development practices. They find that, philosophically and culturally, many development teams resonate with the tenets of Agile development. Our work seems to support this hypothesis, albeit in a different scientific domain, in that the teams involved in our study successfully adopted an Agile framework without much difficulty. 

Our work also proposes a coordination mechanism between separate but dependent Agile projects wherein development of the ``upstream'' project (SCiMMA) must proceed ahead of the ``downstream'' project (SNEWS~2.0) that depends on it. Existing uses of this approach were not found in the literature, but there are several similarly non-traditional implementations of Agile processes involving novel sprint structures. For example, the authors of Ref.~\cite{brenner2015scaled} propose the ``Scaled Agile Framework'' (SAFe) as an additional Agile abstraction to extend Scrum for the coordination of multiple Agile teams. Our work differs in two key ways. First, we articulate a specific methodology based on the existing Agile abstractions of User Stories, Epics, and Backlog. In our case study, the upstream project developed Epics and User Stories that were incorporated into its own Backlog, but the completion of those Backlog tasks would then prompt the creation of User Stories in the downstream Backlog.  Additionally, some User Stories generated by the downstream project were transferred to the Backlog of the upstream project based on their broader relevance to the upstream project's users. Secondly, Ref.~\cite{brenner2015scaled} implicates a more complex structure of organizational dependencies between multiple teams, whereas our case study examines only a single dependency in a bilateral context.

In Ref.~\cite{schuh2017agile}, the authors propose a combination of Waterfall and Agile methods for software development. This is similar to our case study: while we did not formally implement Waterfall methods, the phase-driven approach of our collaboration is comparable to the phase-gate Waterfall process discussed in Ref.~\cite{schuh2017agile}. After each phase, a meeting with leadership members was held in order to assess the status of the project and whether to proceed or halt the collaboration, akin to the ``Go/Kill/Hold/Recycle'' options after each phase-gate. This functioned to steer the outcomes of the Agile development phases without pre-defining the software specifications or development tasks that evolved out of the Agile process, similar to the proposed Project Plan in Ref.~\cite{schuh2017agile}.

In Ref.~\cite{sutherland2005future}, the author proposes a ``MetaScrum'' process in the context of a single Agile development team involving multiple overlapping sprints. While this implementation differs from our case study by considering only one Agile team, it bears similarities in terms of introducing novel Agile processes to coordinate multiple overlapping sprints. Specifically, our case study required coordination and reporting of progress from the downstream sprint (1-week cadence) to the upstream sprint (2-week cadence). While a collaboration-wide MetaScrum meeting was not implemented to manage this coordination, the progress of the downstream project was discussed at meetings involving leadership and management for each organization. In between the collaboration phases, the Agile team met with leadership from each organization to review progress; during phases, the Scrum Master reported progress during SCiMMA sprint reviews and weekly collaboration meetings, and the two Product Owners reported progress at SNEWS monthly team meetings. These processes allowed for broader coordination of the sprints between the organizations, but differed from MetaScrum in that they were not specifically designed with sprint coordination in mind, nor structured to directly solicit feedback from organization leadership.

\section{Conclusion} \label{sec:conclusion}

The development of software to support collaborative science projects requires risk mitigation between the teams involved in order to bridge the chasm between scientific computing and software engineering. This case study describes the risks within a collaborative scientific software development effort and the ways in which the collaboration members confronted them. In this case study, the emergent concerns and needs of each organization were addressed by adopting Agile Scrum practices with an active dialogue between organizations. By working in week-long sprints, the high-level goals of each organization were translated into manageable development tasks. By demonstrating the final product after each phase, the Product Owners could re-evaluate the collaboration's status and the resources required for personnel time and effort; this resulted in enhanced trust in the collaboration. The software developers utilized the practical use-case from the science team to further its software and infrastructure to support the broader MMA community. The science team gained a functional and maintainable prototype of their desired software while also gaining valuable software development experience. The collaboration allowed both organizations to gain useful experience and furthered their science programs in ongoing and future work. The experience also demonstrated several themes of broader interest for the scientific computing community:

\begin{itemize}
    \item Focused, bi-lateral collaboration can streamline development and improve communication of software requirements between users and developers, beyond the traditional open-source pull-request development model. 
    \item The constant communication and incremental tasking promoted by Agile practices allow for software development to be flexible to rapidly changing requirements, which is particularly important when creating scientific applications.
    \item Collaborations between scientists and software developers can accelerate the creation of scientific software applications while also promoting knowledge transfer between domains.
    \item Coordination between separate, but dependent, Agile processes can be done without additional Agile abstractions, but this may require additional planning and communication methods to ensure progress for both teams involved.
\end{itemize}

The framework described herein proved to be beneficial in this study, which involved $\mathcal{O}(10)$ people across two teams and time frames $\mathcal{O}(\mathrm{months})$. Note that these strategies may not generalize to larger-scale or well-established development projects, which may struggle to either implement Agile practices or respond to rapidly-emerging software requirements. However, these practices could still alleviate some of the general difficulties of scientific software engineering, such as interdisciplinary communication, creating sustainable software without relying on external code, and constrained resources and timelines. These emergent and iterative frameworks can additionally support the need for rapid results and ever-changing requirements of the scientist user-group community. This case study indicates that a collaborative development approach centered on an Agile Scrum framework can benefit both the scientific user-group and the software developer team while mitigating some of the difficulties facing the scientific software engineering community.
\section*{Author Contributions}

The alphabetic-order author list reflects a variety of contributions from members in the SCiMMA and SNEWS organizations. Particular contributions to this paper and the SCiMMA-SNEWS collaboration are noted as follows. BC, ALB, and RW made primary contributions to the text. BC and ALB led the demonstrations after each phase of development. BC was the Scrum Master who organized and led the development sprints, supported software development during Phases 1 and 2, and implemented the software on cloud infrastructure. ALB served as a SNEWS product owner, supported software development during Phases 0, 1 and 2, and contributed substantially during development sprints. YX led the SNEWS software development process during Phase 0, 1 and 2. PG led the SCiMMA software development during Phases 1 and 2. Together, PG and YX contributed substantially during development sprints. MWL supported software development during Phases 1 and 2, and implemented the software on prototype infrastructure. AH served as a SNEWS product owner and provided substantial input during development sprints. AB, AH, MJ, RFL, CDT and RW each provided substantial input during progress reviews after each phase. AB, MJ, and CDT proposed the initial collaboration at a National Science Foundation ``Harnessing the Data Revolution'' grant proposal meeting.

The SCiMMA Agile Scrum team (AB, DC, BC, VE, SE, PG, DAH, MJ, WL, CM, SN, DP, RT, CW, and RW) supported core SCiMMA software development prior to and during the collaboration. Additional SCiMMA members (MWGJ, CK) have continued support of the SCiMMA project. The SNEWS Signal Prediction, Alert Formation, and Follow-up working group members (ALB, SYB, WB, MC, AC, MCM, ADO, DD, AGR, SG, AH, RH, SH, JPK, AK, VK, ML, RFL, SL, ML, MWL, JM, DM, RN, EO, BWP, NJ, AR, JR, TS, JCLT, CDT, CFV, CJV, KEW, LW, XJX, YX) supported SNEWS software development prior to and during the collaboration.

\section*{Acknowledgements}

This work is supported by the National Science Foundation ``Windows on the Universe: The Era of Multi-Messenger Astrophysics'' program: ``WoU-MMA: Collaborative Research: A Next-Generation SuperNova Early Warning System for Multimessenger Astronomy'' through grants 1914448, 1914409, 1914447, 1914418, 1914410, 1914416, 1914426; and via the ``HDR-Harnessing the Data Revolution'' program, grant 1940209; and with the ``Office of Advanced Cyberinfrastructure'' through grant 1934752.

\bibliography{bibliography}

\end{document}